%% file: main.tex
\documentclass[conference]{IEEEtran}
\usepackage{grffile}
\usepackage{color}
\usepackage{graphicx}
\usepackage{cite}
\usepackage{multicol}
\usepackage{subcaption}
\usepackage{algcompatible}
\usepackage{algpseudocode}
\usepackage{siunitx}
\usepackage{amsmath}
\usepackage{array}
\usepackage{caption}
\usepackage[linesnumbered,algoruled,boxed,lined]{algorithm2e}
\usepackage{ifthen}
\usepackage{lipsum,graphicx,subcaption}
\usepackage{comment}
\usepackage{romannum}
\usepackage{float}
\usepackage{url}
\usepackage[usestackEOL]{stackengine}
\DeclareMathOperator*{\argmax}{argmax}

\newcommand{\RNum}[1]{\uppercase\expandafter{\romannumeral #1\relax}}

\begin{document}

\title{Balancing Taxi Distribution in A City-Scale Dynamic Ridesharing Service: A Hybrid Solution Based on Demand Learning \\}
\author{\IEEEauthorblockN{Jiyao Li, Vicki H. Allan}
\IEEEauthorblockA{Department of Computer Science, Utah State University, Logan, Utah, 84322\\
Email: jiyao.li@aggiemail.usu.edu, vicki.allan@usu.edu}
}
\maketitle

\begin{abstract}
\input{0_ABSTRACT.tex}
\end{abstract}

\begin{IEEEkeywords}
Smart transportation system, dynamic ridesharing service, correlated pooling, ride-matching, taxi recommender 
\end{IEEEkeywords}

\section{INTRODUCTION}
\input{1_INTRODUCTION.tex}

\section{Related Work}
\label{s2}
\input{2_RELATEDWORK}

\section{System and Performance Metrics}
\label{s3}
\input{3_PROBLEM.tex}

\section{Method}
\label{s4}
\input{4_METHOD.tex}

\section{Experiment}
\label{s5}
\input{5_EXPERIMENT.tex}

\section{Conclusion}
\label{s6}
\input{6_CONCLUSION.tex}

\bibliographystyle{ieeetr}
\bibliography{main}

\end{document}

%% file: 0_ABSTRACT.tex
In this paper, we study the challenging problem of how to balance taxi distribution across a city in a dynamic ridesharing service. First, we introduce the architecture of the dynamic ridesharing system and formally define the performance metrics indicating the efficiency of the system. Then, we propose a hybrid solution involving a series of algorithms: the Correlated Pooling collects correlated rider requests, the Adjacency Ride-Matching based on Demand Learning assigns taxis to riders and balances taxi distribution locally, the Greedy Idle Movement aims to direct taxis without a current assignment to the areas with riders in need of service. In the experiment, we apply city-scale data sets from the city of Chicago and complete a case study analyzing the threshold of correlated rider requests and the average online running time of each algorithm. We also compare our hybrid solution with multiple other methods. The results of our experiment show that our hybrid solution improves customer serving rate without increasing the number of taxis in operation, allows both drivers to earn more and riders to save more per trip, and all with a small increase in calling and extra trip time.

%% file: 1_INTRODUCTION.tex
Dynamic ridesharing services play an important role in smart cities \cite{lai2020review} today. In recent years, there have been a growing number of people participating in dynamic ridesharing services such as Uber and Lyft. A survey of nearly 7000 people in the U.S found that 53 percent of them used a dynamic ridesharing service in 2017, an increase of 15 percent compared to 2016 \cite{ridesharing2018}. In order to strengthen the efficiency of dynamic ridesharing services in such a massive scale market, many ridesharing service companies have begun to study ways to optimize their operations. Most of their research works focus on operation cost reduction, increasing serving rate and rider's satisfaction.

How to balance taxi distribution across the city is a key issue in the dynamic ridesharing business. If taxis across the city are well organized, drivers can save total mileage, increase service rate and rider satisfaction as calling time and waiting time to be picked up can be decreased, and can improve the taxi utilization. However, balancing taxi distribution on a citywide scale is a challenge. First, it is hard to predict the demand of riders across the city at any given time. Rider demand may be high in one place during a certain part of the day but fall dramatically in as little as an hour. Second, the transportation of riders to their destination can also leave taxis in a demand sparse area, leaving many riders unserved even though there are enough taxis for the system. Third, taxis might expend time and fuel to travel to high demand areas of the city, only to find that riders there have been served by others.

Many papers focusing on taxi scheduling in the dynamic ridesharing industry have been published. However, the previous studies suffered from the following limitations: $(\romannum{1})$ Most existing studies proposed different kinds of ride-matching strategies capable of balancing vehicle distribution in a local area to satisfy passenger demand, but failed to schedule vehicles citywide \cite{banerjee2018state} \cite{wang2018deep} \cite{xu2018large} \cite{zhang2017taxi} \cite{li2019ride} ; $(\romannum{2})$ Many research papers designed various methods of guiding empty vehicles to places where drivers had a high chance of picking up passengers. Although vehicles can be scheduled in a wider area, often these systems ignored how to balance supply and demand locally \cite{wen2017rebalancing} \cite{lin2018efficient} \cite{qu2014cost} \cite{jha2018upping}; $(\romannum{3})$ Moreover, while there are also studies that propose price mechanisms to motivate drivers to relocate to places of need, experienced drivers normally do not follow such suggestions \cite{lu2018surge}. 

To address the challenges and limitations of previous works, we propose a hybrid solution based on demand learning, submitting the following contributions:
\begin{itemize}
    \item We apply Value Iteration (VI) \cite{sutton2018reinforcement} to learn the demand pattern using time series from historical data.
    \item We design a novel Correlated Pooling (CP) to group correlated rider requests into clusters.
    \item We design the Adjacency Ride-Matching based on the Demand Learning (ARDL) to balance taxi distribution locally.
    \item Considering the demand pattern difference among areas across the city, we apply a Greedy Idle Movement (GIM) strategy to schedule idle taxis citywide.
\end{itemize}

In Section \ref{s2}, we discuss recent related works. In Section \ref{s3} and \ref{s4}, we introduce the system architecture and performance metrics that estimate the efficiency of the system and propose our hybrid solution. In Section \ref{s5}, we compare our approach to other methods with the city scale real dataset. Conclusions are drawn in Section \ref{s6}.

%% file: 2_RELATEDWORK.tex
Many works have solved supply and demand interaction by ride-matching assignment. \cite{banerjee2018state} proposed an explicit state dependent policy called Scaled MaxWeight (SMW) to balance vehicle distribution in a local area. The policy models a closed queueing network and assigns vehicles from the queue with the most supply. The SMW can achieve exponential decay of the demand dropping probability under the Complete Resource Pooling condition, but it did not consider demand variation in terms of time or day and area of the city.

\cite{wang2018deep} applied deep reinforcement learning to optimize the ride-matching policy. The authors proposed a deep Q-network with action search. A knowledge transfer method was also used to speed up the learning process. The solution adapted to the environment, but there was not any coordination among vehicle agents. \cite{xu2018large} applied historical data to summarize demand and supply patterns by an offline learning step. Each driver-order pair was valued in consideration of the summary in the previous step and dispatch was then solved as a combinatorial optimization problem.

\cite{zhang2017taxi} used Stochastic Gradient Descent (SGD) to predict the acceptance probability of each pair of rider and driver. Ride-matching was solved by a Hill Climbing algorithm. Although both methods can increase serving rate, passengers may suffer from longer waiting time. \cite{li2019ride} proposed a ride-matching strategy based on polar coordinates. This method can improve the platform performance efficiency remarkably, but its performance would be worse when rider request is sparse.

Other works considered scheduling idle vehicles. \cite{wen2017rebalancing} proposed a reinforcement learning approach to move idle vehicles to maintain supply and demand balance. It can reduce the fleet size by $14\%$ while contributing little to overall vehicle distance traveled, but runs the risk of sending so many vehicles into an area of need that supply will be greater than demand. \cite{lin2018efficient} solved this problem by formulating the movement of idle vehicles as a multi-agent problem and contextual multi-agent actor-critic was used. \cite{qu2014cost} formulated a net profit objective function, then used historical data to develop a graph representation of road networks. A novel recursion strategy was applied to find out the optimal route for idle vehicles. \cite{jha2018upping} proposed a Driver Guidance System (DGS) based on street demand prediction. Both authors argued that their methods could maximize drivers' net profit, but they did not show the performance of the system in terms of serving rate.

Some works designed a price mechanism for dynamic ridesharing services. \cite{ma2014real} proposed monetary constraints to encourage riders to participate the dynamic ridesharing service. \cite{kleiner2011mechanism} proposed a second price sealed bid auction for passengers to compete for resources, such that total distance could be minimized and serving rate maximized. \cite{asghari2016price} designed an auction-based approach to distribute demand resources among drivers to maximize drivers' profit. \cite{zheng2019auction} proposed a VCG-like \cite{wooldridge2009introduction} auction mechanism to optimize the overall utility of the auction, while ensuring desirable auction properties such as truthfulness and individual rationality. However, none of these works considered how to motivate drivers to relocate to demand-dense places. \cite{chen2016dynamic} designed a surging price to balance supply and demand, but the efficiency of dispatch vehicles across the city could not be guaranteed.

%% file: 3_PROBLEM.tex
\subsection{Architecture of Dynamic Ridesharing}
\begin{figure}[!h]
\centering
\includegraphics[height=2in]{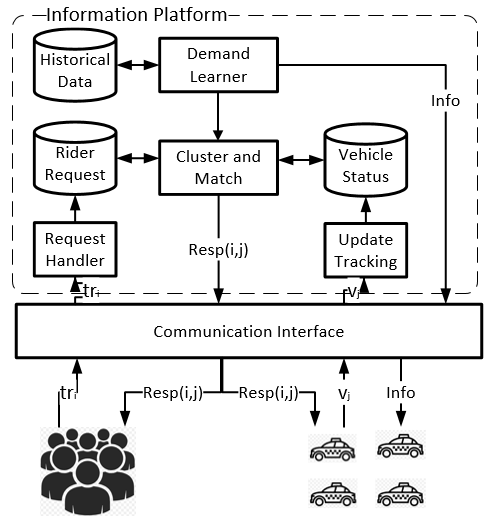}
\caption{Architecture of Dynamic Ridesharing}
\label{DRSys}
\end{figure}

The architecture of a dynamic ridesharing service is shown in Fig.\ref{DRSys}. Each rider $r_i$ can submit a trip request $tr_i$ through smart devices at any time of the day. Each trip request is denoted by a 6-tuple $tr_i=<t,id,o,d,p,k>$, where $t$ is the time of the trip request, $id$ is the trip identification, $o$ and $d$ are pick up and drop off locations, $p$ is patience period, indicating how long the rider will keep requesting the ridesharing service without receiving a response before switching to other alternatives, and $k$ specifies the number of riders in the trip request. For simplicity, we assume that one trip request corresponds to one passenger ($k$=1), but our approach can also be applied in a case where multiple riders share one trip request. After the Information Platform receives the trip request data, it will store them in the Rider Request database until they have expired (exceeded patience period) or have been served.

At the same time, the status of each taxi is tracked in the Information Platform. Each status is denoted by a 5-tuple $v_j=<t,id,l,ca,s>$, where $t$ is current time, $id$ is the taxi identification, $l$ is current location of taxi, $ca$ is current taxi capacity, and $s$ is the route when the taxi has tasks on hand. Otherwise $s$ will be null. After the Information Platform receives a status message, it will update the status records in the Vehicle Status database such that the Information Platform can keep track of each taxi in the system.

The Cluster and Match module is invoked periodically. It clusters rider requests into a group, assigns them to corresponding taxis and sends riders and taxis a response $Resp(i,j)$ notifying them of match results. The Demand Learner learns the demand pattern from the historical data and provides information to the Cluster and Match module as well as to the taxis. 

\subsection{Performance Metrics}
To quantify the efficiency of a solution, we define the following criteria: the first three metrics measure the satisfaction of riders who participate in the dynamic ridesharing service, the fourth and fifth evaluate the average profit of taxi drivers and the total revenue of the Information Platform, and the last three concisely capture the percentage of total riders who can be served , the amount of time period that taxi drivers are serving riders and the poolability at various levels.

\textbf{Metric 1: Calling Time.} The calling time of a trip request $T_r^c$ is the time elapsed between when the trip request is sent by a rider and a response from the Information Platform is received, indicating there is a taxi that can be assigned to the rider, where request $r \in R^+$.

$R^+$ is denoted as the set of all trip requests that have been served and $R^-$ as the set of all trip requests that have been canceled. We define $R=R^++R^-$ where $R$ is the set of all trip requests in the system.

\textbf{Metric 2: Extra Trip Time.} The extra trip time of a trip request $T_r^e$ is the delayed arrival time at the destination for a rider participating in the dynamic ridesharing service, including waiting to be picked up after his or her trip request has been accepted and detour time spent on board, since rider may share taxi with others.

\textbf{Metric 3: Trip Fare.} The trip fare of a trip request $fare(r)$ is the compensation a rider pays for the trip, as shown in Eq.(\ref{fare}), where $F(r)$ is the default price a rider must pay in the solo ride taxi, and $e^{-\lambda T_r^e}$ is the compensation factor.
\begin{equation}
    fare(r) = F(r) \cdot e^{-\lambda T_r^e}
\label{fare}
\end{equation}

Thus, the money saved by a rider for a trip is $F(r)-fare(r)$.

\textbf{Metric 4: Driver Profit} The driver profit $profit(d)$ indicates how much a driver can earn in the dynamic ridesharing service, as shown in Eq.(\ref{profit}), where $S_d$ is the set of riders served by driver $d$, $P_{taxi}$ is the taxi drivers' portion of the trip fare, $C_d$ is the operational cost (e.g. fuel consumption) and $M_c$ is the money spent per unit of operational cost $C_d$.
\begin{equation}
    profit(d) = \sum \limits_{r \in S_d} P_{taxi} \cdot fare(r)-C_d \cdot M_c
\label{profit}
\end{equation}

\textbf{Metric 5: Platform Revenue} The Information Platform shares the trip fares with taxi drivers participating in the dynamic ridesharing service, which is defined as Eq.(\ref{pRevenue}).
\begin{equation}
    plt\_rev = \sum \limits_{r \in R^+} (1-P_{taxi}) \cdot fare(r)
\label{pRevenue}
\end{equation}

\textbf{Metric 6: Serving Rate.} The serving rate reflects the fraction of riders who can be served by taxis, which is defined as Eq.(\ref{serving}), where $|R^+|$ is the number of riders who can be served, and $|R|$ is the number of the total riders.
\begin{equation}
    serving\_rate = \frac{|R^+|}{|R|}
\label{serving}
\end{equation}

\textbf{Metric 7: Taxi Utilization.} Taxi utilization refers to average usage of a taxi over the time period selected for the view, which is defined as Eq.(\ref{utilization}), where $T_d^{idle}$ is the time when a taxi has no delivery task, and $T_{total}$ is the whole time period selected for observation.
\begin{equation}
    taxi\_utilization = \frac{T_{total}-T_d^{idle}}{T_{total}}
\label{utilization}
\end{equation}

\textbf{Metric 8: Poolability.} The $poolability(n)$ is the percentage of riders in a taxi with $n$ riders, as formulated in Eq.(\ref{pool}), where $|R^+|_n$ indicates the number of riders who are in a taxi with $n$ persons for a trip. For example, $|R^+|_1$ is the number of riders who have a taxi alone,  $|R^+|_2$ is the number of riders who are sharing a taxi with one other person.
\begin{equation}
    poolability(n) = \frac{|R^+|_n}{|R|}
\label{pool}
\end{equation}

%% file: 4_METHOD.tex
To serve as many rider requests in various parts of a city as possible, we propose a hybrid solution that can balance supply and demand by scheduling taxis throughout the city. The city is partitioned into various zones. At each zone, we estimate the demand pattern from the current time to the future by the demand learning. Based on the current time demand pattern of each zone, the correlated trip requests are assigned to a taxi by an adjacency ride-matching strategy. We also consider that the most needed taxi zone at each time impacts other zones by the propagation method. A greedy algorithm is applied to schedule taxis when they are in idle mode (no assigned task).  

\subsection{Demand Learning}
The demand learning procedure aims to provide a quantitative understanding of the demand patterns of riders at time t in a specific zone. Given historical data, we build a Markov Reward Process (MRP) \cite{sutton2018reinforcement} with an agent representing each individual zone. The MRP can be represented by a tuple $\langle S,P,R, \gamma \rangle$, where $S$ is a finite set of states and each state is a tuple with timestamp $t_i$ of the day and zone id $z_i$; $P$ is a state transition probability matrix $P_{ss'}=P[S_{t+1}=s'|S_t=s]$, the state transitions are restricted by temporal and spatial properties. For example, state $(t_i, z_j)$ can only transfer to $(t_{i+1}, z_j)$, such that $P_{ss'}=1$; $R$ is a reward function $R_s=E[r_t|S_t=s]$ where symbol $E$ represents the expected value and $r_t$ is the number of rider requests in state s, $\gamma$ is a discount factor that determines the degree of how far the MRP can look into the future and $\gamma \in [0,1]$. 

The value function of the MRP gives the long term evaluation value of state $s$ and can be represented by the formula $V(s) = E[\sum \limits_{k=0}^{+\infty} \gamma ^k R_{t+k} | S_t = s]$, where $S_t$ is the start state. It can also be rewritten in dynamic programming, as shown in Eq.(\ref{dp}), where $S_{t+1}$ is the state of next timestamp:  
\begin{equation}
\begin{split}
    V(s) &= E[\sum \limits_{k=0}^{+\infty} \gamma ^k R_{t+k} | S_t = s] \\
         &= E[R_{t}+\sum \limits_{k=1}^{+\infty} \gamma ^k R_{t+k} | S_t = s] \\
         &= E[R_{t}+\gamma V(S_{t+1})|S_t=s] \\
         &= R_s + \gamma V(S_{t+1})
\end{split}
\label{dp}    
\end{equation}

The learned value function $V(s)$ indicates the degree of rider demand popularity in a specific zone $z_i$ at specific times $t_i$, which is further utilized in the following ride-matching and taxi relocation strategy. Since $P_{ss'}=1$, the next state $S_{t+1}$ at any time is known in advance. Thus, the curse of dimensionality can be avoided. In practice, we perform an implementation based on Value Iteration (VI), as described by Algorithm \ref{MRP}:

\begin{algorithm}
    \SetKwInOut{Input}{Input}
    \SetKwInOut{Output}{Output}
    \Input{Collect historical rider request data $D=(t_i,z_i)$, each state $s_i$ is compose of $(t_i,z_i)$}
    \Output{The value function $V(S)$ of the MRP, where $S$ is the set of $s_i$.}
    \For{$t=T-1$ to $0$}
     {
        \For {each zone $z$ in $Z$}
         {
            Find a subset $D^{(t,z)}$ where $t_i=t$ and $z_i=z$.\\
            Calculate the number of rider requests $R^z_t$ in $z$ at $t$.\\
            \eIf{$t==T-1$}
              {$V(s_i)=R^z_t$}
              {$V(s_i)=R^z_t+ \gamma \cdot V(s_{i+1})$}
         }
     }
    \Return $V(S)$ for all states.
    \caption{Value Iteration (VI) for the MRP.}
    \label{MRP}
\end{algorithm}

\subsection{Correlated Pooling (CP)}
\label{s4_CP}
\textbf{Definition 1: Correlated Trip Request (CTR).} The trip requests $tr_i$ and $tr_j$ are correlated if and only if their trip directions $\overrightarrow{tr_i}$ and $\overrightarrow{tr_j}$ both satisfy the inequality (\ref{CTR}), where $\theta_{CTR}$ is the threshold of CTR.
\begin{equation}
    \operatorname{arcosh} (\frac{\overrightarrow{tr_i}  \cdot \overrightarrow{tr_j}}{|\overrightarrow{tr_i}|  \cdot |\overrightarrow{tr_j}|}) \leq \theta_{CTR}
\label{CTR}
\end{equation}

We can get a vector of trip direction by its source and destination. With trip direction vectors of each trip request in each zone, we compare the vectors to each other and cluster them if the inequality (\ref{CTR}) and taxi capacity can be satisfied, but the complexity of such solution will be $O(kn^2)$, where n is the number of riders in each zone and k is the number of zones. We devise a novel algorithm called Correlated Pooling (CP) based on an index table: first, we create an index table with length $\frac{360}{\theta_{CTR}}$, each key in the index table indicates an angle range. A tuple cluster $tc_k$ is created at each entry of the index table; second, we calculate trip angles by comparing each trip direction with the horizontal line, allowing us to put the trip request into the tuple cluster $tc_k$ of the corresponding entry of the index table; finally, when the total number of trip requests in $tc_k$ fills the capacity of taxi $C_{taxi}$, a new tuple cluster $tc_{k+1}$ will be made for new arrivals. The time complexity of the CP is $O(kn)$. See Algorithm \ref{cluster} for details. 

\begin{algorithm}
    \SetKwInOut{Input}{Input}
    \SetKwInOut{Output}{Output}
    \Input{Trip request list $l^{tr}_{z_i}$ of each zone $z_i$.}
    \Output{An index table $HT_{z_i}$ of each zone $z_i$.}
    Create an index table $HT_{z_i}$ with length $\frac{360}{\theta_{CTR}}$. \\
    Create a tuple cluster $tc_k$ at each entry of $HT_{z_i}$, where $k=0$ for initial. \\
    \For{each zone $z_i$ in $Z$}
     {
        \For{each $tr_j$ in $l^{tr}_{z_i}$}
         {
           Calculate the angle $\theta$ between trip direction $\overrightarrow{tr_j}$ and horizontal line $\overrightarrow{l}$.\\
           Put $tr_j$ into $tc_k$ of $HT_{z_i}[int(\frac{\theta}{\theta_c})]$.\\
           \If{len($tc_k$) = $C_{taxi}$}
            {
              Create another $tc_k$ in $HT_{z_i}[int(\frac{\theta}{\theta_c})]$ for new arrivals, where $k=k+1$.\\
            }
         }
     }
    \Return $HT_{z_i}$ for each zone $z_i$.
    \caption{Correlated Pooling (CP) based on index table.}
    \label{cluster}
\end{algorithm}

\subsection{Adjacency Ride-Matching based on Demand Learning (ARDL)}
\label{s4_ARDL}
\textbf{Definition 2: Supply-Demand Ratio (S-D Ratio).} Given state $s_i$ which is composed of timestamp $t_i$ and zone id $z_i$, the supply-demand ratio $\phi$ is the proportion between the number of available taxis $X(s_i)$ and the value function $V(s_i)$ found by Algorithm \ref{MRP}. It can be defined by the Eq.(\ref{SDR}):
\begin{equation}
    \phi(s_i) = \frac{X(s_i)}{1+V(s_i)}
\label{SDR}
\end{equation}

As we know, riders will look to alternative services if they fail to be assigned a taxi within a reasonable time period. We propose a new assignment policy called Adjacency Ride-Matching based on Demand Learning (ARDL) to balance the distribution of taxis in a local area such that lost demand can be reduced. Assuming that the supplied locations for a rider request are the zone where the request occurs and its adjacent zones, the information platform selects an appropriate taxi from the supplied locations based on the S-D Ratio, which reflects the quantified relationship between the number of taxis and the demand pattern of a specific zone at a specific time. For example, if the S-D Ratio increases, it means that the number of taxis is relatively surplus; if the ratio decreases, the amount of available taxis is not enough for to meet demand. 

Suppose at state $s_i$ where time is $t_i$ and zone is $z_i$, there is a pool of taxis with length $X(s_i)$ and an index table $HT_{z_i}$ with tuple clusters at each zone. Whenever a rider request comes up in a zone, the $\phi$ value of the rider request occurrence zone $z_i$ and its adjacent zones will be calculated and the zone $z_k$ with maximal $\phi$ can be found. A taxi is assigned from zone $z_k$, paired with the tuple cluster of riders, and put into the matching list $l^{M}_{z_i}$ of zone $z_i$. Then the length of the taxi pool of $z_k$ is updated, as described by Algorithm \ref{RMS}:
\begin{algorithm}
    \SetKwInOut{Input}{Input}
    \SetKwInOut{Output}{Output}
    \Input{An index table $HT_{z_i}$ in zone $z_i$ at time $t_i$.}
    \Output{A matching list $l^{M}_{z_i}$ in zone $z_i$ at time $t_i$.}
    \For{each $C_j$ in $HT_{z_i}$}
     {
       $k = \argmax \limits_{i \in O(z_i)} \phi_i(t)$, where $O(z_i)$ is the set of adjacent zones of $z_i$, including $z_i$ itself.\\
       \If{the number of taxis at zone $z_k$ equals to $0$}
        {
            break.\\
        }
       Assign a taxi $T_j$ from zone $z_k$ and bind with $C_j$. \\
       Put $(T_j, C_j)$ into $l^{M}_{z_i}$.\\
       Update taxi amount in zone $z_k$.\\
     }
    \Return $l^{M}_{z_i}$
    \caption{Adjacency Ride-Matching based on Demand Learning (ARDL)}
    \label{RMS}
\end{algorithm}

\subsection{Greedy Idle Movement (GIM)}
\label{s4_GIM}
The ARDL can balance taxis distribution well in a local area wherever there are enough available taxis. In a large city, however, sufficient available taxis cannot be maintained at all times since taxis will deliver riders whose destinations are remote. We also know that taxis will wander around seeking riders while in idle mode. Therefore, directing idle taxis to zones that need taxis can balance taxi distribution across the city.

Simply put, at each unit of time (simulated cycle), the zone with highest number of riders who fail to be served by taxis is selected and its demand pattern (the learned value function) is propagated  to other zones such that a new updated value function can be retrieved across the city. Each taxi in idle mode will calculate the maximized difference between the zone in which it is currently located and the neighboring zones. If the maximized difference exceeds a threshold, then the taxi will move to that adjacent zone. Otherwise it will stay within its current zone.  

The propagation procedure first finds the most needed taxi zone (the zone with the most riders left from last cycle) as base zone $z_b$, then traverses each other zone of the city by Breath First Search (BFS) and updates the original learned value function $V(s_i)$ as $V'(s_i)$, where $s_i=(t_i,z_i)$ and $\alpha \in [0,1]$. The implementation detail is shown in Algorithm.\ref{propagate}.
\begin{algorithm}
    \SetKwInOut{Input}{Input}
    \SetKwInOut{Output}{Output}
    \Input{None}
    \Output{The updated value function $V'(s_i)$ at time $t_i$ }
    Find the zone $z_b$ with most riders left from $t_{i-1}$.\\ 
    Initialize a vector $v$ with length of $|Z|$. \\
    $V'[(t_i,z_b)] = V[(t_i,z_b)]$ \\
    Enqueue $z_b$ into the queue $Q$.\\
    $v[z_b] = true$. \\
    \While{$Q$ is not empty}
     {
         Dequeue a zone $z_q$ from $Q$. \\
         \For{each $z_i$ in Adj($z_q$)}
          {
              \If{$v[z_i] == false$}
               {
                   $V'[(t_i,z_i)] = V[(t_i,z_i)] + \alpha \dot V'[(t_i,z_q)]$\\
                   $v[z_i]=true$\\
                   Enqueue $z_i$ into the queue $Q$.\\
               }
          }
     }
     \Return $V'[s_i]$ 
    \caption{Propagation demand pattern of most needed zone by BFS traverse}
    \label{propagate}
\end{algorithm}

The Greedy Idle Movement (GIM) procedure utilizes the updated value function $V'(s)$ to select the movement direction in the next step. We suppose that taxi $T_j$ located at zone $z_i$ at time $t$, the zone $z_k$ with the maximal $V'(s_k)$ ($s_k=(t,z_k)$) is selected from $O(z_i)$, the adjacent zones of $z_i$. The taxi $T_j$ will move to $z_k$ if the difference between $V'(s_k)$ and $V'(s_i)$ is greater than or equal to a movement threshold $\varphi_m$; otherwise, it will stay in its current zone $z_i$. See Algorithm \ref{gim} for details.      

\begin{algorithm}
    \SetKwInOut{Input}{Input}
    \SetKwInOut{Output}{Output}
    \Input{The current status of a taxi $T_i$}
    \Output{The ID of zone to which a taxi will move}
    \If{$T_i$ is in idle mode}
    {
      //Suppose $T_i$ is located at zone $z_i$ current time $t$.
      $z_k= \argmax \limits_{z_j \in O(z_i)} V'(s_j)$, where $s_j=(t,z_j)$\\
      \eIf{$V'(s_k)-V'(s_i) \geq \varphi_m$}
      {
        //move to $z_k$.\\
        \Return $z_k$
      }
      {
        //stay at current zone $z_i$.\\
        \Return $z_i$
      }
    }
    
    \caption{Greedy Idle Movement (GIM)}
    \label{gim}
\end{algorithm}

%% file: 5_EXPERIMENT.tex
We perform the experiments using the taxi trip requests of the city of Chicago \cite{city2018chicago} between 11:00AM and 11:59PM of a weekday. There are 40,922 requests recorded during the time and each record includes calling timestamp, source zone ID, destination zone ID, and trip fare. We assume that the interval of one simulated cycle $\delta_t$ is 3 minutes. We also assume that if a taxi moves to one of the adjacent zones, the operational cost effort $C_d$ will be 1; in contrast, the $C_d$ will be 0.5 if a taxi just moves around its current zone. The parameters for the dynamic ridesharing service are listed in TABLE \ref{table1}. 

\begin{table}[h]
\captionsetup{skip=0pt}
\caption{Parameter Setting For Dynamic Ridesharing Service} 
\begin{center}
\begin{tabular}{|c|c|c|}
\hline
Notation & Definition & Value\\ \hline
$\delta_t$ & The interval of one simulated cycle & 3 min \\ \hline
$C_d$ & The operational cost for movement& $0.5/1$  \\ \hline
$P_{taxi}$ & Taxi drivers' portion of trip fare & 0.7 \\ \hline
$M_c$ & The money spent per unit of $C_d$ & $\$2$ \\ \hline
$C_{Taxi}$ & The Capacity of a taxi & $4$ \\ \hline
$tr_i.p$ & The time period when a rider keeps calling & 20 min \\ \hline
$\lambda$ & The compensated factor for trip fare & 0.2 \\ \hline
$\varphi_m$ & The idle movement threshold & 0.1 \\ \hline
$\alpha$ & The propagated factor in Algorithm.\ref{gim} & 0.5  \\ \hline
$\gamma$ & The decay factor in Algorithm.\ref{MRP} & 0.8  \\ \hline
\end{tabular}
\end{center}
\label{table1}
\end{table}

We conduct our experiment in two parts. First we conduct a case study analyzing how the variation of the CTR threshold $\theta_{CTR}$ affects the sharing rate, extra trip time, and trip fare of riders. We also study the average online running time of each algorithm. Second, we compare the performance of different integrated solutions as we control various numbers of taxis in the system. To maximize the probability of arriving on time, the stochastic routing technique in \cite{7345570} can be used to select the optimal path among points in the delivery route. All the simulations are implemented by Python 3.5 and executed by a machine with Intel Corei7-3770CPU (3.4GHz, quad-core) and 16GB memory.

\subsection{A Case Study}
\begin{figure}[!h]
  \centering
  \includegraphics[width=0.7\linewidth]{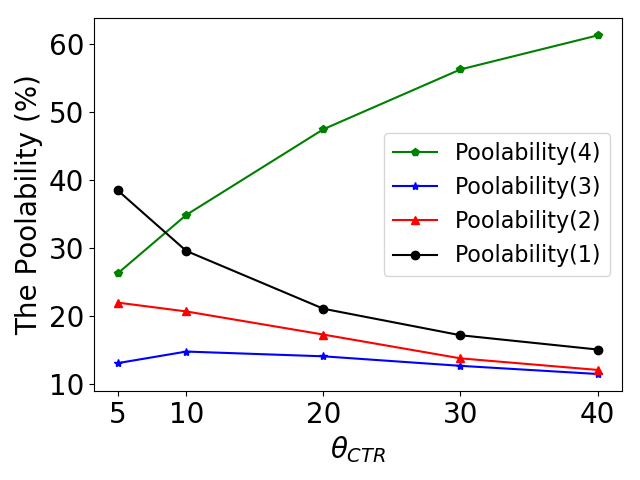}\quad
  \caption{The Poolability}
  \label{case1}
\end{figure}

First, we study the poolability along with the variation of the $\theta_{CTR}$, as shown in Fig.\ref{case1}. We observe that as the $\theta_{CTR}$ increases, the Poolability(4) surges up to $61\%$ while the portion of riders taking a taxi alone (the Poolability(1)) decreases dramatically to $15\%$. This is because the number of riders of each Poolability(n) (except 4 that is the taxi capacity) tend to increase as the $\theta_{CTR}$ increased. For example, in Fig.\ref{case1}, riders at the Poolability(1) have more space to upgrade to higher levels, and that is why the curve of the Poolability(1) decreases dramatically; in contrast, the Poolability(4) receives riders from the lower levels as the $\theta_{CTR}$ increases, which is why it increases significantly. We may also observe that the curves of the Poolability(2) and Poolability(3) remain stable all around the $\theta_{CTR}$ variation. This is because the additional riders from the Poolability(1) compensate for the riders lost to the Poolability(4).

\begin{figure}[!h]
  \centering
  \includegraphics[width=0.7\linewidth]{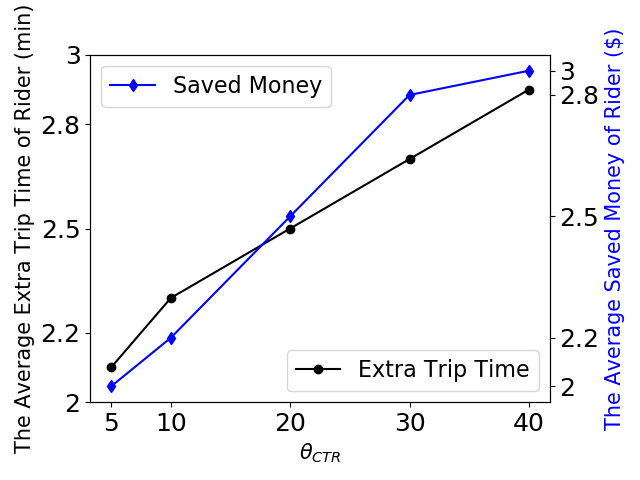}\quad
  \caption{Extra Trip Time $\&$ Saved Trip Fare}
  \label{case2}
\end{figure}

Fig.\ref{case2} indicates that the extra trip time and trip fare increase as the $\theta_{CTR}$ goes up. From Fig.\ref{case2}, riders tend to be pooled together in one taxi as the $\theta_{CTR}$ relax. This will also lead to more extra trip time on board for riders who can save more money compensating for the extra trip time. 

Considering the extra trip time is not too long (about 2.5min) and saved trip fare of riders is nearly the maximum achieved (nearly $3$) when $\theta_{CTR}=30$, we set $\theta_{CTR}$ as $30$. In additional, the average online running time of each algorithm is listed in TABLE \ref{table2}.
\begin{table}[ht]
\centering
\caption{The Average Online Time of Each Algorithm}
\label{table2}
\begin{tabular}[t]{lcc}
\hline
&\Centerstack{Average Online Running Time of \\ Each Simulated Cycle (sec)}\\
\hline
Correlated Pooling (CP) &0.006\\
\Centerstack{Adjacency Ride-Matching based \\ on Demand Learning (ARDL)} &0.034\\
Greedy Idle Movement (GIM) &0.05\\
\hline
\end{tabular}
\end{table}

\subsection{Result Comparisons}
For evaluation purposes, we implemented the SMW policy from \cite{banerjee2018state}. We also integrated three different hybrid solutions with the series of algorithms proposed in \ref{s4}. Detailed descriptions of each solutions are presented below:
\begin{itemize}
    \item \textbf{SMW:} Taxis are in the closed queues that were modeled at beginning. Whenever a rider request comes up, a taxi will be assigned from the queue with maximized weight.
    \item \textbf{SMW+CP:} Rider requests are pooled into groups by Correlated Pooling (CP) described in \ref{s4_CP}, then the SMW is applied to assign taxis to each group.
    \item \textbf{ARDL+CP:} Rider requests are grouped by CP, then the Adjacency Ride-Matching Demand Learning (ARDL) described in \ref{s4_ARDL} is applied to pair rider groups and taxis. 
    \item \textbf{ARDL+CP+GIM:} Rider requests are grouped by CP, and ARDL is used to match riders and taxis. Whenever a taxi has no task on hand (idle mode), it will use Greedy Idle Movement (GIM) described in \ref{s4_GIM} to seek for the route to the zones in need of taxis.
\end{itemize}

\begin{figure}[!h]
  \centering
  \includegraphics[width=0.7\linewidth]{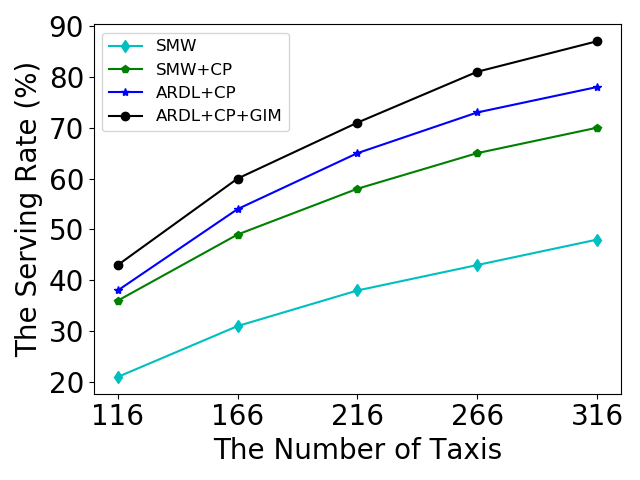}\quad
  \caption{The Serving Rate}
  \label{SR}
\end{figure}

\begin{figure}[!h]
  \centering
  \includegraphics[width=0.7\linewidth]{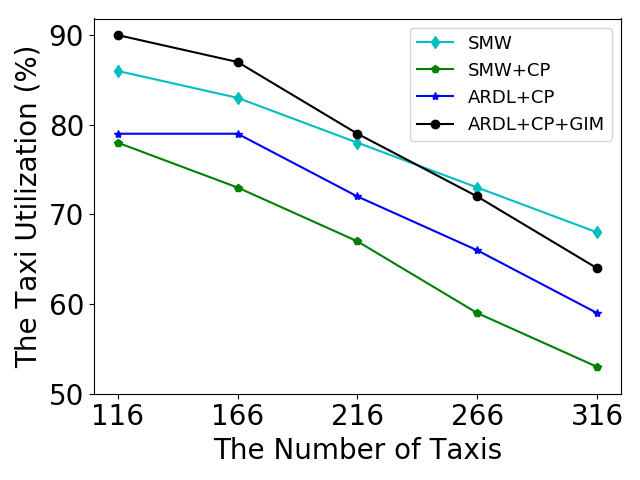}\quad
  \caption{The Taxi Utilization Rate}
  \label{TUR}
\end{figure}

Fig.\ref{SR} and Fig.\ref{TUR} present the performance metrics of the serving rate and taxi utilization. We observe that the serving rate improves while the taxi utilization rate decreases as the number of taxis goes up, indicating that more taxis participating in dynamic ridesharing can help more riders get service and reduce drivers working hours. Fig.\ref{SR} compares each hybrid solution in terms of the serving rate, we discovers that CP, ARDL and GIM has positive effect on the serving rate since SMW+CP$>$SMW, ARDL+CP$>$SMW+CP and ARDL+CP+GIM$>$ARDL+CP.

Fig.\ref{TUR} shows the taxi utilization in each hybrid solution. We observe that CP has a negative effect on utilization rate, since one taxi can serve multiple riders such that the taxi's working hours decrease, but the hybrid solution ARDL+CP+GIM performs well because the taxis will still be directed to the areas of greatest need when they have no task on hand.

\begin{figure}[!h]
  \centering
  \includegraphics[width=0.7\linewidth]{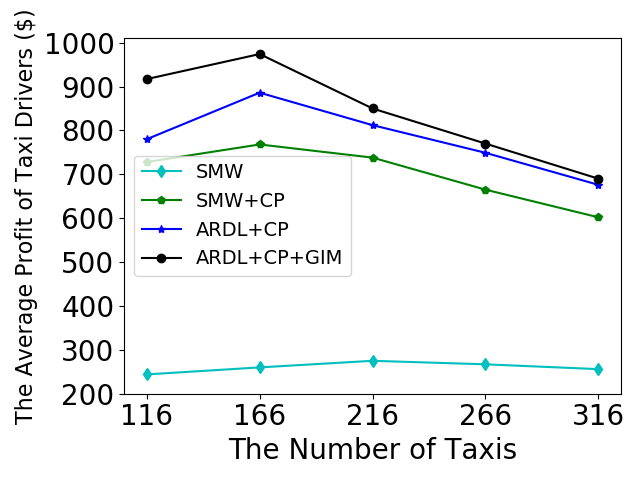}\quad
  \caption{The Average Profit of Taxi Drivers}
  \label{dprofit}
\end{figure}

\begin{figure}[!h]
  \centering
  \includegraphics[width=0.7\linewidth]{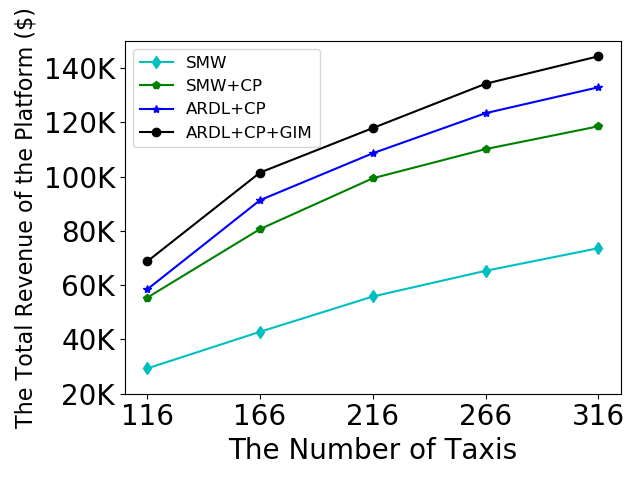}\quad
  \caption{The Total Revenue of Information Platform}
  \label{pltrevenue}
\end{figure}

Fig.\ref{dprofit} and Fig.\ref{pltrevenue} reflect the economic aspect of the system. Fig.\ref{dprofit} plots the average profit of taxi drivers of each hybrid solution. The profit of SMW+CP is greater than SMW since CP can dramatically increase the serving rate and therefore the amount of revenue. The ARDL is better than SMW in terms of average profit since ARDL+CP$>$SMW+CP. The hybrid solution ARDL+CP+GIM can earn more profit than the other three solutions since the GIM can motivate idle taxis to places where riders fail to be served, although the operational cost $C_d$ increases for idle movement of taxis.

On the other hand, Fig.\ref{pltrevenue} shows the total revenue of the Information Platform. Eq.(\ref{pRevenue}) shows that the total revenue of the platform is proportionate to the total trip fare with a strong correlation between it and the serving rate. In the figure, the total revenue of the hybrid solution ARDL+CP+GIM is higher than the other three methods, along with various numbers of taxis, because the serving rate of the ARDL+CP+GIM is also highest among all solutions. 

\begin{figure}[!h]
  \centering
  \includegraphics[width=0.7\linewidth]{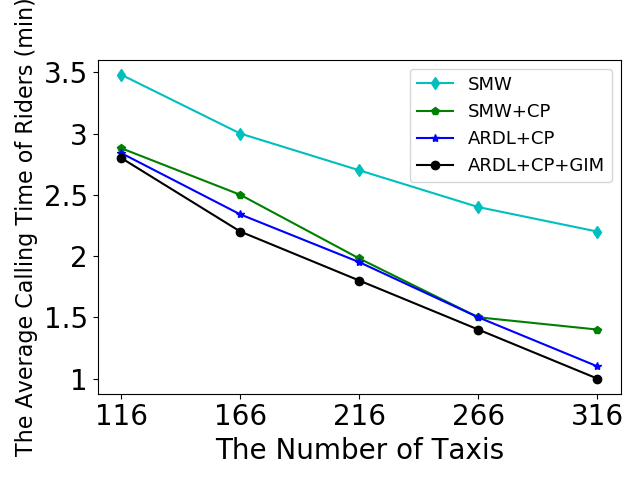}\quad
  \caption{The Average Calling Time of Riders}
  \label{PM3}
\end{figure}

Fig.\ref{PM3} presents the average calling time of riders of each hybrid solution. The calling time is the time between when a rider send a request and a taxi is assigned to him or her. Considering that taxis can pick up riders in a short period, the assignment policy will assign taxis in areas adjacent to the riders. If we want to minimize the calling time, taxis should be distributed strategically across the city. Ideally, a taxi will be nearby wherever and whenever a rider request is submitted. In the figure, we find that the calling time of the hybrid solution ARDL+CP+GIM is the lowest and ARDL+CP is the second lowest. This is because the ARDL can schedule taxis to the areas of need locally while the GIM can distribute taxis efficiently citywide.  

%% file: 6_CONCLUSION.tex
In our study of effective citywide taxi distribution in the dynamic ridesharing industry, we examined the architecture of dynamic ridesharing services and defined the performance metrics that estimate the efficiency of these systems. We then proposed a series of algorithms: Correlated Pooling (CP), Adjacency Ride-matching based on Demand Learning (ARDL) and Greedy Idle Movement (GIM). To test their effectiveness, we applied city-scale real world trip request data. The results show that $316$ taxis may serve up to $90\%$ of riders, a taxi driver can earn about $\$53$ per hour, while a rider can save nearly $\$3$ for a trip, the information platform can potentially receive a revenue of $\$144,412$ totally, and the cost in time is only a $0.6$ min wait for the response from the information platform as well as $2.6$ min of extra trip time.

In future, we will apply reinforcement learning in the taxi idle movement.